\documentstyle[seceq,epsf]{ptptex}
\title{%        %You can use \\ for explicit line-break
Analyses of Type Ia Supernova Data \\
in Cosmological Models with a Local Void
}
\author{
Kenji {\sc Tomita}\footnote{E-mail address: tomita@yukawa.kyoto-u.ac.jp
} 
}

\inst{
Yukawa Institute for Theoretical Physics, 
Kyoto University, Kyoto 606-8502}

\recdate{August 3, 2001      %Editorial Office will fill in this.
}

\abst{%       %this abstract is neglected when [addenda] or [errata]
The data for type Ia supernovae obtained by the High-z SN Search Team
and Supernova Cosmology Project are analyzed using inhomogeneous
cosmological models with a local void on scales of about 200 Mpc, to
derive the best-fit values of model parameters and the confidence
contours. The $\chi^2$ fitting is found to be slightly better than
that in homogeneous models. It is shown that (1) the best-fit values are
most sensitive to the ratio $R$ of the outer Hubble constant, $H_0^{\rm
II}$, to the inner Hubble constant, $H_0^{\rm I}$, (2) the best-fit outer 
density parameter, $\Omega_0^{\rm II}$, and cosmological constant
parameter, $\lambda_0^{\rm II}$, are, respectively, increasing and
decreasing functions of $R$, and (3) $(\Omega_0^{\rm II}, \lambda_0^{\rm
II})$ can be $(1, 0)$ for $R \approx 0.8$.
Moreover, it is shown that these models are naturally consistent
with the new supernova data (SN1997ff) with $z = 1.7$.
}

\begin{document}

\maketitle

%ch1
\section{Introduction}

In present day cosmology, the most important observations are those of 
the [magnitude $m$ - redshift $z$]
relation for type Ia supernovae (SNIa) and the CMB anisotropy. In the
former observations, SNIa play the role of the best standard candles, and two 
groups, the High-$z$ SN Search Team\cite{sch,riessa,riessb} and the
Supernova Cosmology Project,\cite{perl} have to this time observed 50 
and 60 SNIa, respectively. Their results suggest that a significant 
amount of dark energy fills our
universe, under the assumption of its homogeneity and isotropy.

Inhomogeneous models with a local void on scales of about 200 Mpc have 
been studied by the present 
author\cite{tma} in connection with puzzling behavior of cosmic bulk 
flows\cite{hud,will} and it was subsequently shown that it may be
possible to explain the accelerating behavior of high-$z$ SNIa in 
[$m , z$] relation without a cosmological constant.\cite{tmb} A
historical survey of works concerned with local voids is given in
a previous work.\cite{tmc} Because these inhomogeneous
models include many parameters (such as the inner and outer values of
the Hubble constant and the density parameter, the cosmological constant, and
the boundary radius), we recently examined the dependence of the
above relation on these parameters, in comparison with the relations
in homogeneous models with $(\Omega_0, \lambda_0) = (0.3, 0.7)$ and
$(0.3, 0.0)$.\cite{tmd}

Observationally, recent galactic redshift 
surveys\cite{mari,marz,folk,zucc} show that in the region approximately 
$200$ -- $300 h^{-1}$ Mpc from us, the distribution of galaxies may be 
inhomogeneous. Moreover, a large-scale inhomogeneity suggesting the 
existence of a wall around the void on scales of $\sim 250 h^{-1}$ Mpc
has recently been found by Blanton et al.\cite{blant} from the SDSS 
commissioning data.  Similar walls on scales of $\sim 250 h^{-1}$ Mpc
have been found from the Las Campanas and 2dF redshift surveys
near the Northern and Southern Galactic Caps.\cite{schect,folk,cole} 
These results may imply that there is a local void of radius  
$\sim 200$ -- $300 h^{-1}$ Mpc in which we live. 

In this paper we analyze directly the SNIa data of the above-mentioned
 two groups using our inhomogeneous models with a local void on scales 
 of about 200
Mpc, and obtain the confidence contours as well as the best-fit values
of the model parameters. In \S 2 we give the basic formulation of the
distance modulus and the $\chi$ square for statistical fitting.  In 
\S 3 we derive the best-fit values of model parameters and the confidence
contours. In addition, we compare the behavior of our models with the new
data for $z = 1.7$,\cite{new} and we find that they appear to be naturally
consistent with it. In \S 4, discussion and concluding remarks are given. 

%ch2
\section[]{Distance modulus and the $\chi$ square}

The theoretical distance modulus is defined by
\begin{equation}
  \label{eq:m1}
\mu_0^p = 5 \log \Big(d_{\rm L}/ {\rm Mpc}\Big) + 25,
\end{equation}
where $d_{\rm L}$ is the luminosity distance, related to the
angular-diameter distance $d_{\rm A}$ by
\begin{equation}
  \label{eq:m2}
d_{\rm L} = (1 + z)^2 d_{\rm A},
\end{equation}
along the light ray to a source S with redshift $z$. This distance 
modulus is compared with the observed one, given by
\begin{equation}
  \label{eq:m3}
\mu_0 = m_B - M_B,
\end{equation}
where $M_B$ and $m_B$ are the peak absolute magnitude and the
corresponding apparent magnitude of a standard SNIa in the B band,
respectively. 

In this paper we consider spherical inhomogeneous cosmological models
which consist of an inner homogeneous region, V$^{\rm I}$, and an
outer homogeneous region, V$^{\rm II}$, with a boundary of radius
$\sim 200$ Mpc. The observer's position O deviates generally from the
center C, but is assumed to be close to C, compared with the
boundary. For the
off-center observer, the angular-diameter distance $d_{\rm A}$ depends 
not only on $z$, but also on the angle between the vectors 
$\overline{\rm CO}$ and $\overline{\rm CS}$, where S is the source. 
In our previous papers,\cite{tma,tmb} the behavior of distances
for off-center observers was studied. Since the angular dependence is
small for remote sources, however, we can ignore it for simplicity,
assuming that we are approximately at the center. Then $d_{\rm A}$
depends on the source redshift $z$, the inner and outer Hubble
constants $H_0^{\rm I}$ and  $H_0^{\rm II}$, the inner and outer
density parameters $\Omega_0^{\rm I}$ and  $\Omega_0^{\rm II}$, the
outer $\Lambda$ parameter $\lambda_0^{\rm II}$, and the boundary
redshift $z_1$. The inner $\Lambda$ parameter $\lambda_0^{\rm I}$ is
related to $\lambda_0^{\rm II}$ by $\lambda_0^{\rm I} = \lambda_0^{\rm
II} (H_0^{\rm II}/H_0^{\rm I})^2$. The equations to be solved to 
derive $d_{\rm A}$  and the junction conditions are 
Eqs. (5) -- (11) in Ref.~10).
  
The best-fit values for the cosmological parameters are determined
using the $\chi$-square expression
\begin{eqnarray}
  \label{eq:m4}
\chi^2 = &\sum_i & \Big[\mu^p_{0,i} (z_i| H_0^{\rm I}, H_0^{\rm II},
\Omega_0^{\rm I}, \Omega_0^{\rm II}, \lambda_0^{\rm II}, z_1) -
\mu_{0,i}\Big]^2 \cr
&& / (\sigma^2_{\mu 0,i} + \sigma^2_{mz,i}),  
\end{eqnarray}
where $\sigma_{\mu 0}$ is the measurement error of the distance
modulus and $\sigma_{mz}$ is the dispersion in the distance modulus
corresponding to the dispersion of the galaxy redshift $\sigma_z$ (coming
from peculiar velocity and uncertainty). $\sigma_{mz}$ is related to
$\sigma_z$ as
\begin{equation}
  \label{eq:m5}
\sigma_{mz} = {5 \over \ln 10} \Big({1 \over d_{\rm L}}{\partial
d_{\rm L} \over \partial z}\Big).
\end{equation}

Next, the probability distribution function (PDF) must be obtained to
derive the confidence contours and the most likely values of the
model parameters. It is expressed as
\begin{equation}
  \label{eq:m6}
p(H_0^{\rm I}, H_0^{\rm II}, \Omega_0^{\rm I}, \Omega_0^{\rm II},
\lambda_0^{\rm II}, z_1 |\mu_0) \propto \exp \Big(- {1 \over 2} \chi^2\Big), 
\end{equation}
which is consistent with Eq.~(9) in Ref.~2).

In the present inhomogeneous models there are six parameters, and it is too 
complicated to consider all of them simultaneously. For this reason,
 we treat here 
several cases with specific values of $z_1, H_0^{\rm II}/H_0^{\rm I},$ 
and $\Omega_0^{\rm I}$. Then, the corresponding normalized PDF can be
expressed as
\begin{eqnarray}
  \label{eq:m7}
&&p(H_0^{\rm I}, \Omega_0^{\rm II}, \lambda_0^{\rm II} |\mu_0)\cr
&& = {\exp
(- {1 \over 2} \chi^2) \over \int^\infty_{-\infty} d H_0^{\rm I} 
\int^\infty_{-\infty} d \lambda_0^{\rm II}  \int^\infty_{\Omega_{0l}}
\exp (- {1 \over 2} \chi^2) d  \Omega_0^{\rm II}}
\end{eqnarray}
for given $z_1, H_0^{\rm II}/H_0^{\rm I},$ and $\Omega_0^{\rm I}$,
where $\mu_0$ is the set of distance moduli used and $\Omega_{0l}$ is
the assumed lower limit of $\Omega_0^{\rm II}$, being $0$ or $-
\infty$. Physically, the region in which $\Omega_0^{\rm II} < 0$ is
meaningless, although it is significant statistically. To obtain a 
probability distribution independent of $H_0^{\rm I}$, we can consider 
the following quantity:
\begin{equation}
  \label{eq:m8}
p(\Omega_0^{\rm II}, \lambda_0^{\rm II} |\mu_0) =
\int^\infty_{-\infty} p(H_0^{\rm I}, \Omega_0^{\rm II}, \lambda_0^{\rm
II} |\mu_0) d H_0^{\rm I}.
\end{equation}
The contours are determined by $p(\Omega_0^{\rm II}, \lambda_0^{\rm II}
|\mu_0)$.

\begin{figure}
\epsfxsize = 10cm
\centerline{\epsfbox{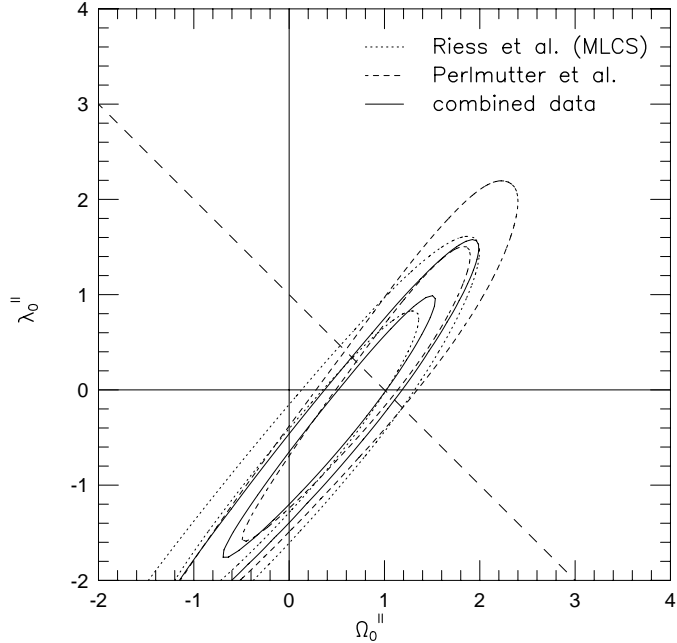}}
\caption{The 68.3 \% and 95.4 \% confidence contours in the $\Omega_0^{\rm
II}$ - $\lambda_0^{\rm II}$ plane in the case  $(z_1, H_0^{\rm II}/
H_0^{\rm I}, \Omega_0^{\rm I}) = (0.080, 0.82, {\rm A})$. 
The dotted curves represent the data of Riess et al. (1998), the dashed
curves those of Perlmutter et al. (1999), and the solid curves
 the combined data.
  \label{fig:1}}
\end{figure}
\begin{figure}
\epsfxsize = 10cm
\centerline{\epsfbox{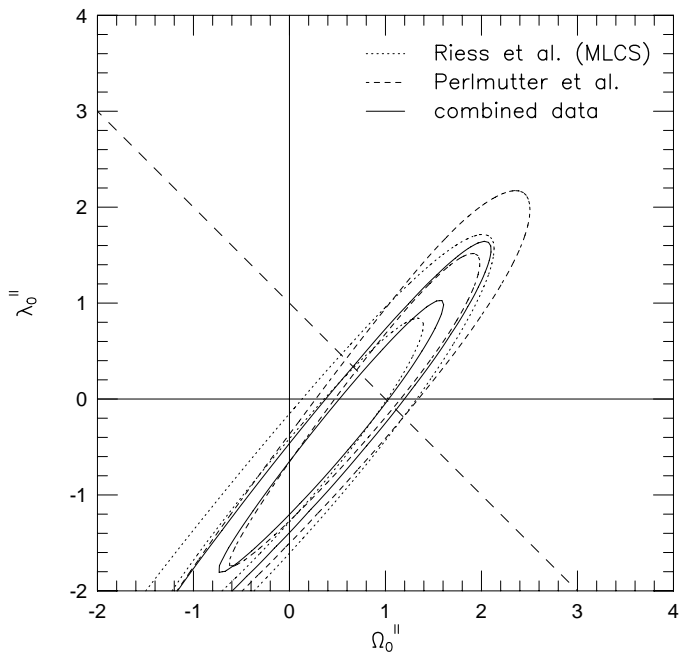}}
\caption{The 68.3 \% and 95.4 \% confidence contours in the $\Omega_0^{\rm
II}$ - $\lambda_0^{\rm II}$ plane in the case  $(z_1, H_0^{\rm II}/
H_0^{\rm I}, \Omega_0^{\rm I}) = (0.080, 0.82, {\rm B})$. 
  \label{fig:2}}
\end{figure}
\begin{figure}
\epsfxsize = 10cm
\centerline{\epsfbox{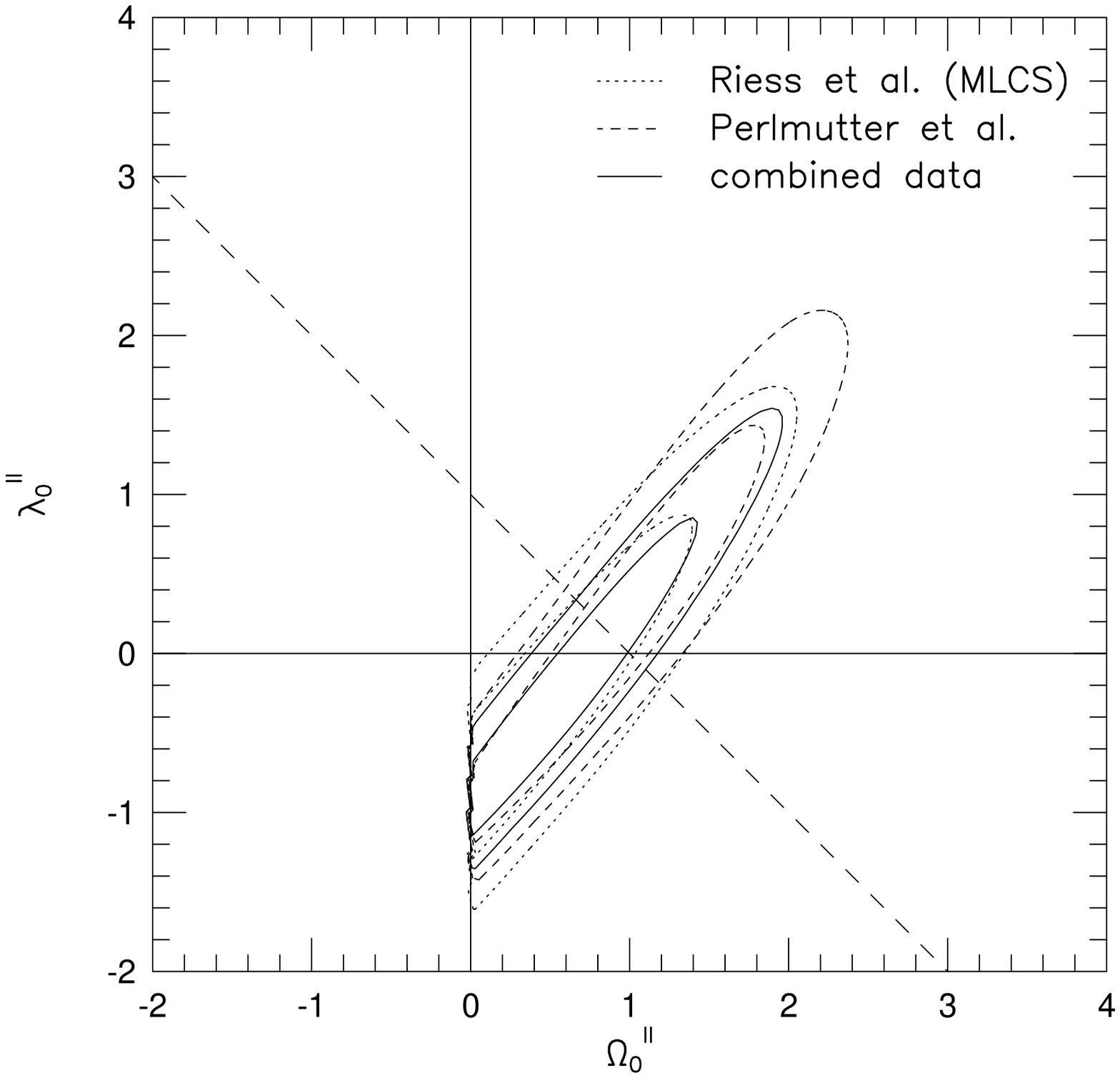}}
\caption{The 68.3 \% and 95.4 \% confidence contours in the $\Omega_0^{\rm
II}$ - $\lambda_0^{\rm II}$ plane in the case  $(z_1, H_0^{\rm II}/
H_0^{\rm I}, \Omega_0^{\rm I}) = (0.080, 0.82, {\rm A})$ with
$\Omega_{0l} = 0$.
  \label{fig:3}}
\end{figure}

%ch3
\section{Supernova data and the fitting}
At present, we have two SNIa data sets to be used for determining
cosmological parameters, that of HSST (Schmidt et al.\cite{sch}, Riess et
al.\cite{riessa}) with 50 SNIa and that of SCP (Perlmutter et
al.\cite{perl}) with 60 SNIa. 18 SNIa are common to both data.
In the former, there are two types of $\mu_0$ data, that due to the
Multicolor Light Curve Shape method (MLCS) and that due to the 
template-fitting method. 
In each method, we can use the data for $z_i, \mu_{0,i}$ and
$\sigma_{\mu0,i}$ (with $i = 1$ -- $50$) given in their tables. 
Following Riess et al.\cite{riessa} for $\sigma_z$, we adopt $\sigma_z =
200$ km s$^{-1}$ and add $2500$ km s$^{-1}$ in quadrature to
$\sigma_z$ for SNIa whose redshifts were determined from broad
features in the SN spectrum.

In Ref.~4), the data for $m^{\rm eff}_{B,i}, z_i,
\sigma_{mz,i}$ and $\sigma_{z,i} \ (i = 1$ -- $60)$ are given, but the
values of $\mu_{0,i}$ were not published and are not available. As 
shown by Wang,\cite{wang} however, we have for 18 SNIa observed by both
teams
\begin{equation}
  \label{eq:m9}
M_B^{\rm MLCS} \equiv m_B^{\rm eff} - \mu_0^{\rm MLCS} = - 19.33 \pm 0.25,
\end{equation}
where $m_B^{\rm eff}$ is the effective B-band magnitude of SNIa given
by Perlmutter et al., and $\mu_0^{\rm MLCS}$ is the corresponding data 
for $\mu_0$ due to MLCS method in Riess et al.\cite{riessa} 

For the corrected B-band peak absolute magnitude derived by Hamuy et
al.,\cite{hamm} on the other hand, we have
\begin{equation}
  \label{eq:m10}
M_B^{\rm MLCS} - M_B^{\rm H96} = -0.047 \pm 0.270,
\end{equation}
which is sufficiently small, compared with the counterpart due to the
template-fitting method. Accordingly, we adopt the MLCS data of
Riess et al.'s two data sets, and derive the values of $\mu_0$ from
the data set of Perlmutter et al. using the relation
\begin{equation}
  \label{eq:m11}
\mu_0 = m_B^{\rm eff} - \bar{M}_B \ {\rm with} \ \bar{M}_B = - 19.33,
\end{equation}
following Wang.\cite{wang}

From the two data sets we make a combined data set consisting of 50
SNIa (from Riess et al.) and 42 SNIa (from Perlmutter et al.), in
which we adopt the data of Riess et al. for 18 common data (due to Wang's
procedure\cite{wang}). 

The above three kinds of data were compared with the theoretical models 
using the following conditions concerning $z_1, H_0^{\rm II}/H_0^{\rm I},$ 
and $\Omega_0^{\rm I}$. For $z_1$, we mainly used $z_1 = 0.080$,
corresponding to the radius $240/ h^{\rm I}$ Mpc, where $H_0^{\rm I} = 
100 h^{\rm I}$ km s$^{-1}$ Mpc$^{-1}$. We compare the PDF in this case
with the PDF in cases with $z_1 = 0.067~(cz_1/H_0^{\rm I} = 
200/ h^{\rm I}$ Mpc) and $z_1 = 0.100~(cz_1/H_0^{\rm I} = 
300 / h^{\rm I}$ Mpc) below.
For $H_0^{\rm II}/H_0^{\rm I},$ we consider three cases, with $H_0^{\rm
II}/H_0^{\rm I} = 0.80, 0.82$ and $0.87$ (or $H_0^{\rm I}/H_0^{\rm II} = 
1.25, 1.20$ and $1.15$), respectively.

For $\Omega_0^{\rm I}$, we first consider the inner low-density case
(A), in which
\begin{equation}
  \label{eq:m12}
\Omega_0^{\rm I} = 0.3 \quad {\rm for} \ \Omega_0^{\rm II} > 0.6  
\end{equation}
and 
\begin{equation}
  \label{eq:m13}
\Omega_0^{\rm I} = \Omega_0^{\rm II}/2 \quad {\rm for} \ 
\Omega_0^{\rm II} < 0.6.  
\end{equation}
For comparison, we also consider the equi-density case (B) with 
\begin{equation}
  \label{eq:m14}
\Omega_0^{\rm I} = \Omega_0^{\rm II} \ \Big(H_0^{\rm II}/H_0^{\rm I}\Big)^2.
\end{equation}
Since $\rho_0^j \propto (H_0^j)^2 \Omega_0^j \quad (j =$ I, II), we have 
$\rho_0^{\rm I} = \rho_0^{\rm II}$ in this case.

\begin{figure}
\epsfxsize = 10cm
\centerline{\epsfbox{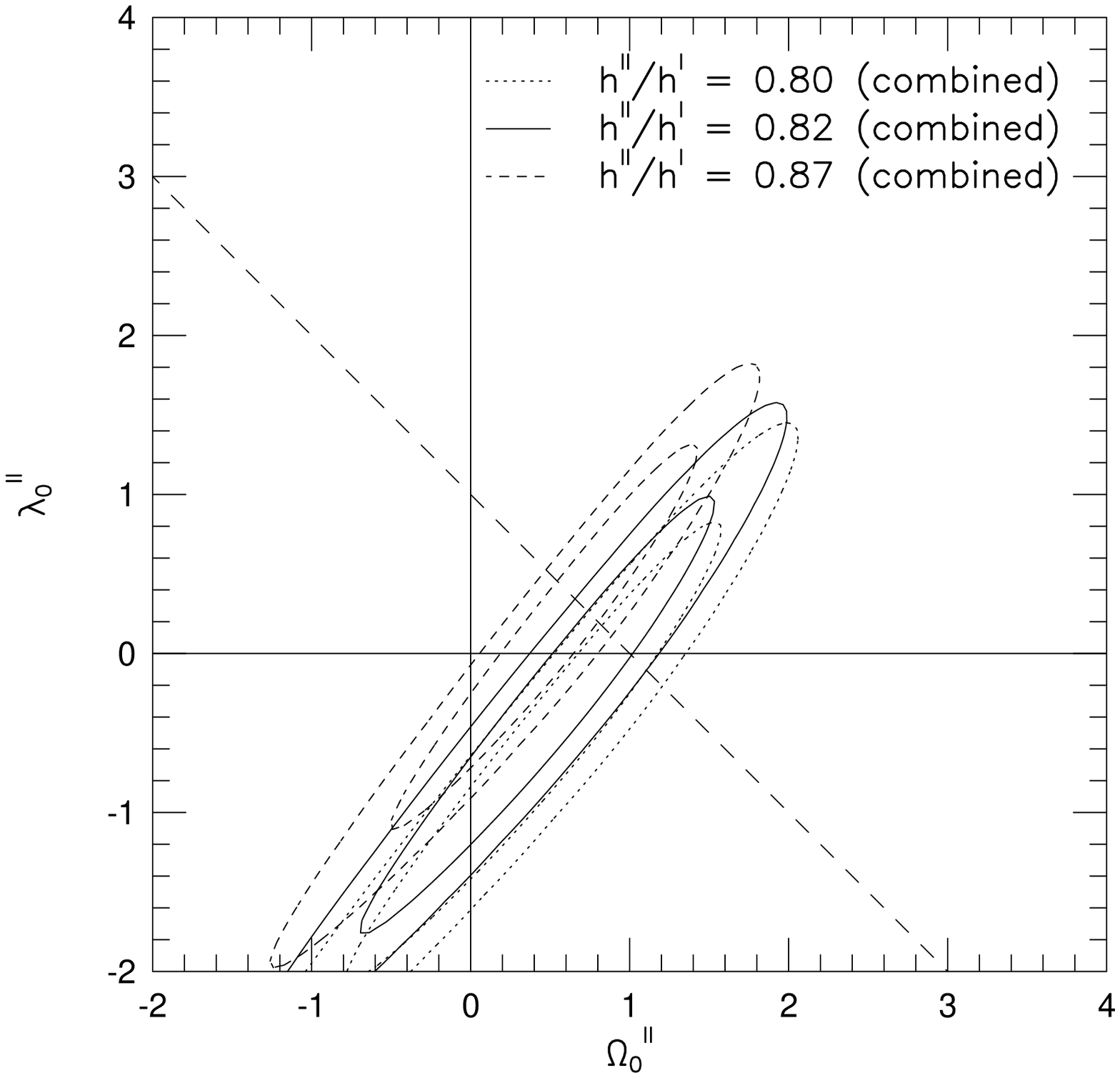}}
\caption{The 68.3 \% and 95.4 \% confidence contours in the $\Omega_0^{\rm
II}$ - $\lambda_0^{\rm II}$ plane for the combined data. The dotted
 curves, solid curves and dashed curves represent the cases $H_0^{\rm
 II}/H_0^{\rm I} = 0.80, 0.82,$ and $0.87$, respectively.
  \label{fig:4}}
\end{figure}
\begin{figure}
\epsfxsize = 10cm
\centerline{\epsfbox{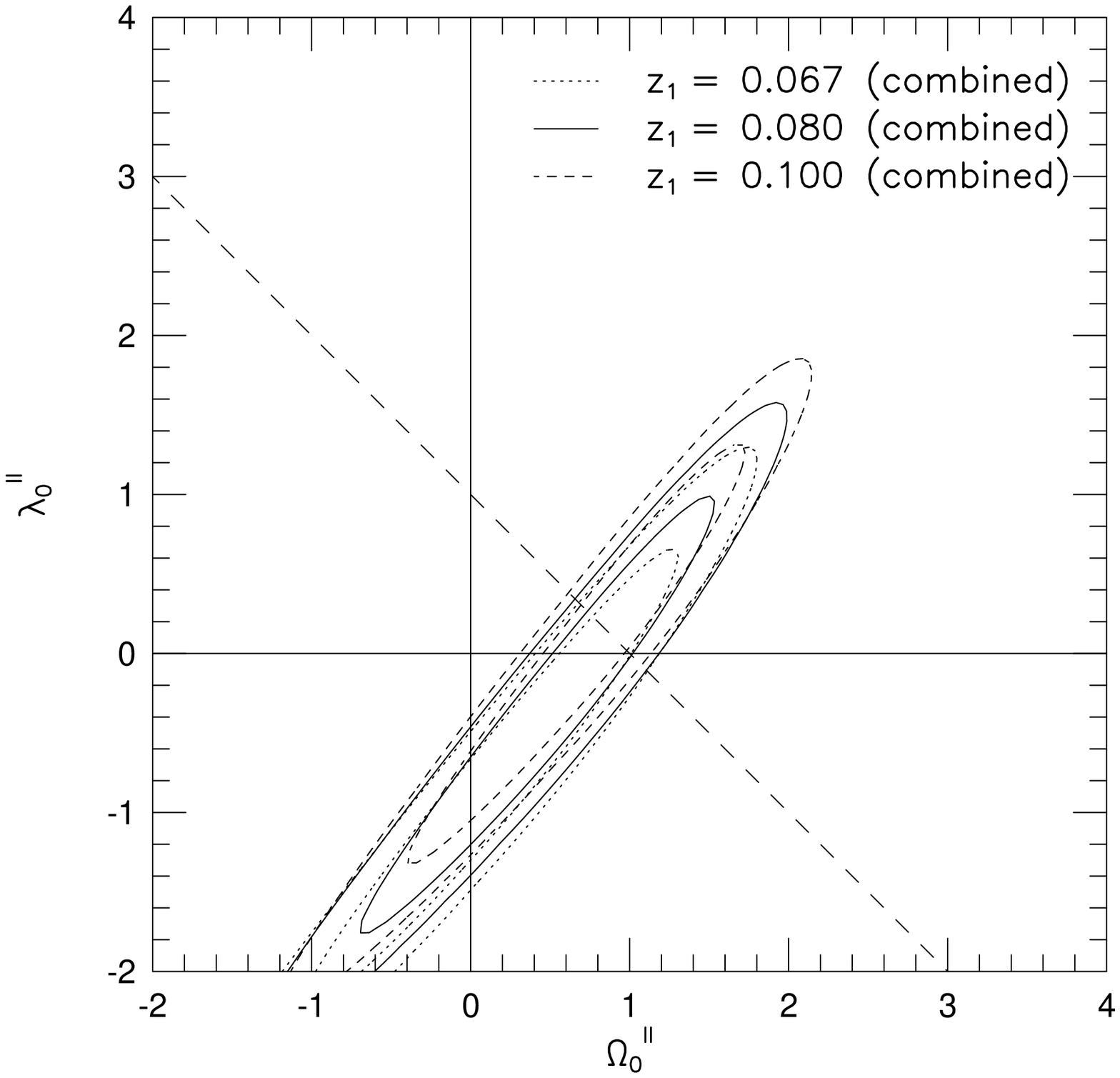}}
\caption{The 68.3 \% and 95.4 \% confidence contours in the $\Omega_0^{\rm
II}$ - $\lambda_0^{\rm II}$ plane for the combined data. The dotted
 curves, solid curves and dashed curves represent the cases $z_1 =
 0.067, 0.080,$ and $0.100$, respectively.
  \label{fig:5}}
\end{figure}

In Figures 1 -- 5, we plot the 68.3 \% and 95.4 \%
confidence contours in the $\Omega_0^{\rm II}$ - $\lambda_0^{\rm II}$
plane. In Figs. 1, 2 and 3, we use three kinds of data and treat the
three cases $(z_1, H_0^{\rm II}/H_0^{\rm I}, \Omega_0^{\rm
I}) = (0.080,\ 0.82, {\rm A}), \ (0.080, \ 0.82, {\rm B}),$ \ and
$(0.080, 0.82, {\rm A})$, with $\Omega_{0l} = 0$, respectively. 
Only in the case 
$\Omega_{0l} = 0$, the region of $\Omega_0^{\rm II} < 0$ is excluded. 
It is found comparing these figures that the difference between the two 
cases A and B is small,
and the difference between the two cases with $\Omega_{0l} =
-\infty$ and $0$ is also small.

In Figs. 4 and 5, we display the difference between the contours 
in the cases $R \equiv H_0^{\rm II}/H_0^{\rm I} = 0.80, \ 0.82$ 
and $0.87$ and $z_1 =
0.067, \ 0.080$ and $0.100$, respectively, using the combined data.
It is interesting that as $R$ decreases, the contours move in the
direction of increasing $\Omega_0^{\rm II}$ and decreasing $\lambda_0^{\rm
II}$, and for $R = 0.80$, the flat 
case with vanishing cosmological constant (the Einstein-de Sitter
model) is at the center of the contours.

% ---------------------------------
\begin{table}
\centering
\caption{Model parameters determined for the homogeneous models. Here,
$\chi_\nu^2$ is the value of $\chi^2$ per degree of freedom. 
$H_0 = 100h$ km s$^{-1}$ Mpc$^{-1}$. The errors on $h$ expressed here are
only statistical.}
\label{tab:1(homog)}
\begin{tabular}{|c|c|c|c|} \hline
%{data} & \multicolumn{1}{c}{Riess et al.}
%&\multicolumn{1}{c}{Perlmutter et al.} & 
%\multicolumn{1}{c}{combined} \\ \hline
{data} & {Riess et al.}
&{Perlmutter et al.} & {combined} \\ \hline
$h$ & $0.65\pm 0.01$ & $0.666 \pm 0.02$ & $0.65 \pm 0.01$ \\ 
$\Omega_0$ & $0.2 \pm 0.6$ & $1.0 \pm 0.4$ & $0.7 \pm 0.4$ \\
$\lambda_0$ & $0.7 \pm 0.8$ & $1.7 \pm 0.6$ & $1.2 \pm 0.5$\\
$\chi_\nu^2$ & $1.11$ & $1.57$ & $1.45$\\ \hline
%$(\Omega_0)_{\rm flat}$ & $0.9 \pm 0.2$ & $0.9\pm0.2$ & $0.9\pm0.1$\\
%  \hline
\end{tabular}
\end{table}
\begin{table}
\centering
\caption{Model parameters determined in the case $(z_1, H_0^{\rm II}/
H_0^{\rm I}, \Omega_0^{\rm I}) = (0.080, 0.80, {\rm A})$. 
$\chi_{\nu}^2$ is the value of $\chi^2$ per degree of freedom. 
$H_0^{\rm I} = 100h^{\rm I}$ km s$^{-1}$ Mpc$^{-1}$. 
The errors on $h^{\rm I}$  are only statistical.}
\label{tab:2(80A)}
\begin{tabular}{|c|c|c|c|} \hline
%{data} & \multicolumn{1}{c}{Riess et al.}
%&\multicolumn{1}{c}{Perlmutter et al.} & 
%\multicolumn{1}{c}{combined} \\ \hline
{data} & {Riess et al.}
&{Perlmutter et al.} & {combined} \\ \hline
$h^{\rm I}$ & $0.64\pm 0.01$ & $0.64 \pm 0.02$ & $0.64 \pm 0.01$ \\ 
$\Omega_0^{\rm II}$ & $0.1 \pm 0.7$ & $0.9 \pm 0.6$ & $0.5 \pm 0.5$ \\
$\lambda_0^{\rm II}$ & $-0.9 \pm 1.0$ & $-0.1 \pm 0.8$ & $-0.5 \pm 0.7$\\
$\chi_{\nu}^2$ & $1.05$ & $1.62$ & $1.42$\\ \hline
$(\Omega_0^{\rm II})_{\rm flat}$ & $1.0 \pm 0.2$ & $1.0\pm0.2$ & $1.0\pm0.1$\\
  \hline
\end{tabular}
\end{table}
\begin{table}
\centering
\caption{Model parameters determined in the case $(z_1, H_0^{\rm II}/
H_0^{\rm I}, \Omega_0^{\rm I}) = (0.080, 0.82, {\rm A})$. 
$\chi_{\nu}^2$ is the value of $\chi^2$ per degree of freedom. 
$H_0^{\rm I} = 100h^{\rm I}$ km s$^{-1}$ Mpc$^{-1}$. 
The errors on $h^{\rm I}$ are
only statistical.}
\label{tab:3(82A)}
\begin{tabular}{|c|c|c|c|} \hline
%{data} & \multicolumn{1}{c}{Riess et al.}
%&\multicolumn{1}{c}{Perlmutter et al.} & 
%\multicolumn{1}{c}{combined} \\ \hline
{data} & {Riess et al.}
&{Perlmutter et al.} & {combined} \\ \hline
$h^{\rm I}$ & $0.64\pm 0.01$ & $0.64 \pm 0.02$ & $0.64 \pm 0.01$ \\ 
$\Omega_0^{\rm II}$ & $0.1 \pm 0.7$ & $0.9 \pm 0.7$ & $0.6 \pm 0.5$ \\
$\lambda_0^{\rm II}$ & $-0.7 \pm 1.1$ & $0.1 \pm 1.0$ & $-0.2 \pm 0.7$\\
$\chi_{\nu}^2$ & $1.05$ & $1.61$ & $1.42$\\ \hline
$(\Omega_0^{\rm II})_{\rm flat}$ & $0.9 \pm 0.2$ & $0.9\pm0.2$ & $0.9\pm0.1$\\
  \hline
\end{tabular}
\end{table}

In Table I, the best-fit values of the cosmological parameters in the
homogeneous models are listed for later comparison. In Tables II, III and
IV, we display the best-fit values of $h^{\rm I}, \Omega_0^{\rm II}$ and
$\lambda_0^{\rm II}$ with $1 \sigma$ error bars, and the 
corresponding values of $\chi_\nu^2$ in the cases 
$H_0^{\rm II}/H_0^{\rm I} = 0.82, \ 0.87$ and $0.80$, respectively,
assuming $z_1 = 0.080$ and $\Omega_0^{\rm I}$ (A). 
The error bars on $h^{\rm I}$ include only statistical errors, not 
the contributions from the errors of the absolute magnitudes. The
values of $\chi_\nu^2$ for the data of Perlmutter et al. is 
found to be much larger than that for the data of 
Riess et al.  This results from the situation that for Perlmutter et
al. we made
artificially the data for $\mu_0$, using the average absolute magnitude.
In the data of Riess et al., $\chi_\nu^2$ in our inhomogeneous
models is smaller than that in the homogeneous models (Table 1), and
so these data seem to be fitted better by our inhomogeneous models 
or the inhomogeneity
of Hubble constant, though more parameters are included in our case.
 To the bottom of 
Tables II -- V, we add $(\Omega_0^{\rm II})_{\rm flat}$, the density
parameter in the case that the outer region is spatially flat. 
 
In Table V we list the best-fit values in the case of  $\Omega_0^{\rm
I}$ (B), assuming $H_0^{\rm II}/H_0^{\rm I} = 0.82, z_1 =
0.080$. From a comparison of Tables II and V, we see that the best-fit
values are not sensitive to $\Omega_0^{\rm I}$.
In Table VI, moreover, we list the values for the cases with 
$z_1 = 0.067$ and $0.100$, 
assuming $H_0^{\rm II}/H_0^{\rm I} = 0.82$ and $\Omega_0^{\rm I}$ (A).
It is found from this table that, because of the smallest
$\chi_{\nu}^2$, the fitting for $z_1 = 0.080$ is
better than that in the cases with $z_1 = 0.067$ and $0.100$. This
result may be connected with the fact that the boundary radius of
the observed local void is likely to be between $200/h^{\rm I}$ Mpc
and $300/h^{\rm I}$ Mpc.

Finally, we compare the behavior of our models with the new supernova
data  for $z = 1.7$.\cite{new} In Fig. 6 we show the $\Delta (m - M)$ - 
$\log z$ diagram, where $\Delta (m - M)$ is the difference between \ $m - M$ 
\ in each model and that in the empty homogeneous model $(\Omega_0 =
\lambda_0 = 0)$. Here we have adopted two models with a local void in
which 
$(\Omega_0^{\rm II}, \lambda_0^{\rm II}) = (1.0, 0.0)$ and $(0.6,
0.0)$. In both models we use $\Omega_0^{\rm I} = 0.3, \ \lambda_0^{\rm I} 
= 0$ and $z_1 = 0.067$ as the other parameters, to which the behavior 
of the models is not sensitive. The upper, middle and lower curves 
in each model correspond to $R \ (\equiv
H_0^{\rm II}/H_0^{\rm I}) = 0.80, 0.82$ and $0.87$, respectively.
For comparison, we also display the behavior of three homogeneous models with
$(\Omega_0, \lambda_0) = (0.35, 0.65), (0.35, 0.0)$ and $(1.0, 0.0)$,
and depict in the diagram the binned data $(z < 1)$ and the new data 
(SN 1997ff) for $z = 1.7$, which are given in Fig. 11 of Ref.~18).
  It is found from Fig. 6 that 
our models can be consistent with the new data, in contrast to the flat 
homogeneous model with $(\Omega_0, \lambda_0) = (0.35,
0.65)$. In particular, the model with $(\Omega_0^{\rm II}, \lambda_0^{\rm
II}) = (1.0, 0.0)$, which is the  Einstein-de Sitter model in the
external region, seems to naturally accord with all data with $R = 0.80 
$ and $0.82$. In the model with $(\Omega_0^{\rm II}, \lambda_0^{\rm
II}) = (0.6, 0.0)$, the case $R = 0.87$ is best among the three cases.
If more data for $z > 1$ are provided, we shall be able to distinguish 
more clearly the behavior of the models with a local void and that of the
homogeneous models with dominant cosmological constant or dark energy.

\begin{figure}
\epsfxsize = 12cm
\centerline{\epsfbox{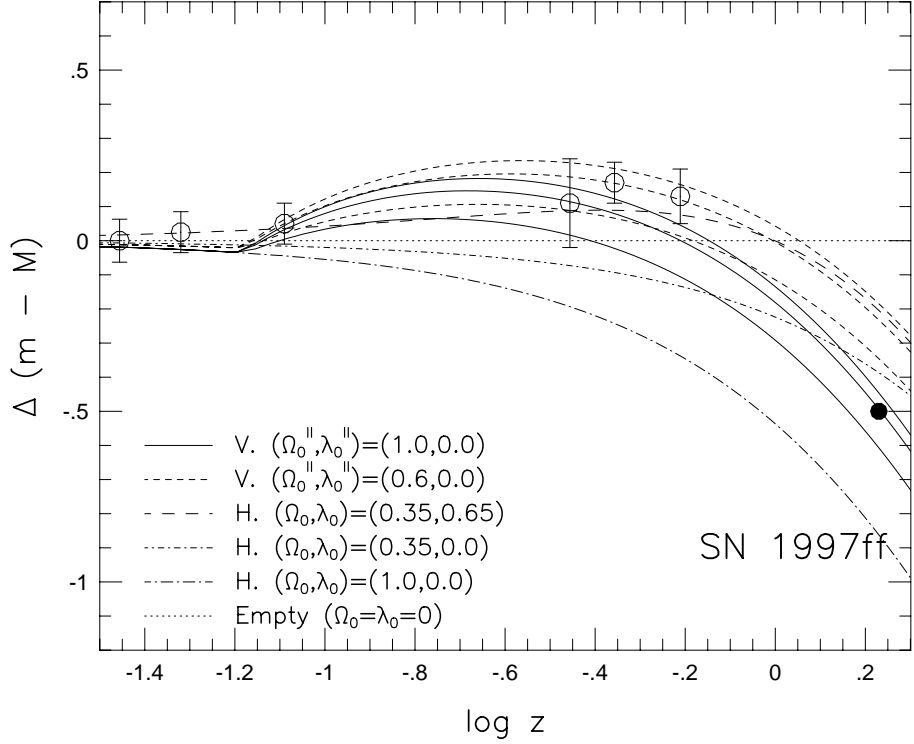}}
\caption{The $\Delta (m - M)$ - $\log z$ diagram for models with a local void 
(V) and homogeneous models (H). Here $\Delta (m - M) \equiv (m - M)$ 
in each model $- \ (m - M)$ in the empty model $(\Omega_0 = \lambda_0 =
0)$. In the two V models, we have upper, middle and lower curves
corresponding to $R \ (\equiv H_0^{\rm II}/H_0^{\rm I}) = 0.80, 0.82$
and $0.87$, respectively.  For comparison the SNIa binned data $(z <
1)$ and the new data for SN 1997ff $(z = 1.7)$ are also depicted in the diagram.
 \label{fig:6}}
\end{figure}
%

%ch4
\section{Concluding remarks}
In this paper we derived confidence contours and best-fit parameters
for inhomogeneous models with a local void using two sets of SNIa data
and the combined data, and we found that (1) they are very sensitive to the
ratio $R$ of the outer Hubble constant $H_0^{\rm
II}$ to the inner Hubble constant $H_0^{\rm I}$, \ (2) the best-fit outer 
density parameter $\Omega_0^{\rm II}$ increases and the cosmological constant
parameter $\lambda_0^{\rm II}$ decreases as functions of $R$, and 
(3) $(\Omega_0^{\rm II}, \lambda_0^{\rm
II})$ can be $(1, 0)$ for $R \approx 0.8$.
It is thus found that the existence of our local void may solve the puzzling
{\it cosmological-constant} problem.
 
However, we ignored the directional dependence 
of the $[m, z]$ relation, which may be an important factor, especially for
nearby SNIa, since magnitudes of SNIa are measured by off-center
observers. If the data of angular positions of observed SNIa are
published, it should be possible to take into account the directional 
dependence, and the fitting may be improved. 

The flux averaging proposed by Wang\cite{wang} was not carried out
 here, but
it may be important when many high-$z$ data with $z > 1.0$ appear,
because they would be much affected by the lensing effect.  

From a comparison with the new data for $z = 1.7$, we found that our
models are naturally cosnsistent with it, in contrast to the
homogeneous models with acceleration due to the cosmological constant or
dark energy. For $z > 1$, the behavior of curves represents the
deceleration of the universe, which depends strongly on the equation of
state of the constituent matter. In our models with a local void, this is
mainly pressureless matter,
while the homogeneous models with the dominant cosmological constant and
hypothetical dark energy have negative pressure comparable with the
mass energy. More data with $z > 1.5$ is needed
 in order to distinguish these two types of models more 
clearly and remove the fluctuations due to the lensing effect.

\begin{table}
\centering
\caption{Model parameters determined for the case $(z_1, H_0^{\rm II}/
H_0^{\rm I}, \Omega_0^{\rm I}) = (0.080, 0.87, {\rm A})$. 
$\chi_{\nu}^2$ is the value of $\chi^2$ per degree of freedom. 
$H_0^{\rm I} = 100
h^{\rm I}$ km s$^{-1}$ Mpc$^{-1}$. The errors on $h^{\rm I}$ are
only statistical.}
\label{tab:4(87A)}
\begin{tabular}{|c|c|c|c|} \hline
%{data} & \multicolumn{1}{c}{Riess et al.}
%&\multicolumn{1}{c}{Perlmutter et al.} & 
%\multicolumn{1}{c}{combined} \\ \hline
{data} & {Riess et al.}
&{Perlmutter et al.} & {combined} \\ \hline
$h^{\rm I}$ & $0.64\pm 0.01$ & $0.65 \pm 0.02$ & $0.64 \pm 0.01$ \\ 
$\Omega_0^{\rm II}$ & $0.2 \pm 1.0$ & $0.7 \pm 0.7$ & $0.6 \pm 0.6$ \\
$\lambda_0^{\rm II}$ & $-0.2 \pm 1.1$ & $0.7 \pm 1.1$ & $0.3 \pm 0.8$\\
$\chi_{\nu}^2$ & $1.05$ & $1.61$ & $1.42$\\ \hline
$(\Omega_0^{\rm II})_{\rm flat}$ & $0.7 \pm 0.2$ & $0.7\pm0.2$ & $0.7\pm0.1$\\
  \hline
\end{tabular}
\end{table}
\begin{table}
\centering
\caption{Model parameters determined for the case $(z_1, H_0^{\rm II}/
H_0^{\rm I}, \Omega_0^{\rm I}) = (0.080, 0.82, {\rm B})$. 
$\chi_{\nu}^2$ is the value of $\chi^2$ per degree of freedom. 
$H_0^{\rm I} = 100h^{\rm I}$ km s$^{-1}$ Mpc$^{-1}$. 
The errors on $h^{\rm I}$ are only statistical.}
\label{tab:5(82B)}
\begin{tabular}{|c|c|c|c|} \hline
%{data} & \multicolumn{1}{c}{Riess et al.}
%&\multicolumn{1}{c}{Perlmutter et al.} & 
{data} & {Riess et al.}
&{Perlmutter et al.} & {combined} \\ \hline
%\multicolumn{1}{c}{combined} \\ \hline
$h^{\rm I}$ & $0.64\pm 0.01$ & $0.64 \pm 0.02$ & $0.64 \pm 0.01$ \\ 
$\Omega_0^{\rm II}$ & $0.0 \pm 0.7$ & $0.8 \pm 0.7$ & $0.5 \pm 0.5$ \\
$\lambda_0^{\rm II}$ & $-0.8 \pm 1.1$ & $0.0 \pm 1.0$ & $-0.3 \pm 0.8$\\
$\chi_{\nu}^2$ & $1.05$ & $1.62$ & $1.42$\\ \hline
$(\Omega_0^{\rm II})_{\rm flat}$ & $0.9 \pm 0.2$ & $0.9\pm0.2$ & $0.9\pm0.1$\\
  \hline
\end{tabular}
\end{table}
\begin{table}
\centering
\caption{Model parameters determined for the case $(z_1, H_0^{\rm II}/
H_0^{\rm I}, \Omega_0^{\rm I}) = (0.080, 0.82, {\rm A})$. 
$\chi_{\nu}^2$ is the value of $\chi^2$ per degree of freedom. 
$H_0^{\rm I} = 100h^{\rm I}$ km s$^{-1}$ Mpc$^{-1}$. 
The errors on $h^{\rm I}$ are only statistical.}
\label{tab:6(82z1)}
\begin{tabular}{|c|c|c|c|c|} \hline
%{data} & \multicolumn{1}{c}{Riess et al.}
%&\multicolumn{1}{c}{Riess et al.} & 
%\multicolumn{1}{c}{combined}& \multicolumn{1}{c}{combined} \\ \hline
{data} & {Riess et al.} & {Riess et al.}
&{combined} & {combined} \\ \hline
$z_1$ & 0.067 & 0.100& 0.067& 0.100 \\
$h^{\rm I}$ & $0.64\pm 0.01$ & $0.64 \pm 0.01$ & $0.64 \pm 0.01$ &
$0.64 \pm 0.01$\\  
$\Omega_0^{\rm II}$ & $-0.3 \pm 0.7$ & $0.4 \pm 0.7$ & $0.2 \pm 0.5$&
$0.9 \pm 0.5$ \\ 
$\lambda_0^{\rm II}$ & $-1.3 \pm 1.1$ & $-0.3 \pm 1.1$ & $-0.7 \pm
0.8$& $0.2 \pm 0.8$\\ 
$\chi_{\nu}^2$ & $1.10$ & $1.13$ & $1.45$& $1.46$\\ \hline
\end{tabular}
\end{table}

\section*{Acknowledgements}
The author is grateful to referees for helpful comments.
 This work was supported by a Grant-in Aid for Scientific Research 
(No.~12440063) from the Ministry of Education, Science, Sports and
Culture, Japan. He owes also to the YITP computer system for the
 numerical analyses.

%\newpage
%ref

\label{lastpage}

\end{document}